
\magnification 1200
\def\sqr#1#2{{\vcenter{\vbox{\hrule height.#2pt
     \hbox{\vrule width.#2pt height#1pt \kern#1pt
     \vrule width.#2pt}
     \hrule height.#2pt}}}}
\def\square{\mathchoice\sqr34\sqr34\sqr{2.1}3\sqr{1.5}3}
\def\ah{\rm arctanh}
\centerline{\bf The Energy-Momentum Tensor in Fulling-Rindler Vacuum}
\bigskip
\centerline{Renaud Parentani\footnote{*}{on leave from: Universit\'e
Libre de Bruxelles, Service de Physique Th\'eorique. Campus Plaine, C.P. 225,
Bd. du Triomphe, B1050 Belgium}}
\bigskip
\centerline{The Racah Institute of Physics, The Hebrew University of
Jerusalem, Jerusalem 91904, ISRAEL}
\bigskip
\bigskip
\bigskip
\bigskip

{\parindent = 12 pt \narrower\narrower\narrower
{\bf Abstract.} The energy density in Fulling-Rindler vacuum, which is known to
be negative "everywhere" is shown to be positive and singular on the
horizons in such a fashion as to guarantee the positivity of the total
energy. The mechanism of compensation is displayed in detail.\par}
\vskip 4.5truein
ULB preprint ULB-TH-15/92
\vfill
\eject

\noindent
{\bf{1. Introduction}}

In connection with the problem of the radiation emitted and absorbed by an
accelerating observer [1, 2, 3, 10, 14], an interesting conceptual problem has
arisen. What is the behavior of the energy density in Fulling-Rindler (FR)
vacuum?

We remind the reader that Minkovski vacuum presents itself as a thermal
distribution of FR quanta within the quadrant containing the accelerating
trajectory [1, 4, 5]. One often describes excitation of the detector in
terms of absorbtion of these FR quanta so the above question naturally arises
(i.e. what would the state look like when all FR quanta are absorbed?).

Existent calculations [6, 7] in the literature give the result that the
energy density in FR vacuum is negative. This has been interpreted as the
absence of the thermal FR quanta which characterize Minkovski vacuum. When
this density is integrated over all space, the total energy in FR vacuum
would be therefore less than that of Minkovski vacuum. But this later is the
ground state.

In this paper, this dilemma is solved. We show that the energy density is
singular and positive in such a fashion that the total (integrated) energy
is positive.

Our method proceeds through the introduction of an ultra violet regulator
($\varepsilon$) which permits the evaluation of the energy density close
and on the horizon. As $\varepsilon \to 0$, the energy density tends to a
distribution. For all points off the horizon, this density is the negative
value previously obtained. Yet its integral is positive.

We note in passing that the same behaviour exists in Boulware vacuum near
the Schwarzschild horizon [6, 7], as well as near the horizons of the static
patch in de Sitter space, in the state of vacuum defined by the static
system.

\medskip
\noindent
{\bf 2. The FR Vacuum and Minkovski Vacuum.}

For simplicity we work in two space time dimensions with a massless scalar
field
$\phi$. It is then particularly useful to work in light cone coordinates to
benefit from the conformal invariance of this massless theory. First we
introduce $U$ and $V$ defined by:

$$\eqalign{U&=t - z\cr V&=t + z \cr}\eqno(1)$$

\noindent
where $t$ and $z$ are the usual Minkovski coordinates. We define also $u_R$ and
$v_R$ (and $u_L, v_L$) by:

$$\eqalign{u_R &=  -\theta (-U) {1 \over a} ln (-aU)\cr
v_R &= \theta (V) {1 \over a} ln (aV)\cr}\eqno(2)$$

$$\eqalign{u_L &= \theta(U) {1 \over a} ln (aU)\cr
v_L &= -\theta(-V) {1 \over a} ln (-aV)\cr}\eqno(3)$$

\noindent
where $a$ has the dimension of a [length]$^{-1}$ and where $R(L)$
indicates that only the right (left) hand side of the Minkovski plane is
covered by those coordinates.

In the first quadrant (i.e. $|t| < z > 0$), $u_R$ and $v_R$ are related to the
usual Rindler coordinates ($\tau, \rho$) by:

$$\eqalign{{(u_R +v_R) / 2} &= \tau =~(1/a){\ah}({t / z})\cr
e^{(v_R - u_R)} &= a^2 \rho^2  = a^2(z^2 - t^2)\cr}\eqno(4)$$

\noindent
$\tau$ is the proper time of a system following the trajectory $\rho = 1 / a$.

Let us now obtain the FR and the usual Minkovski vacua through analysis of
the solutions of $\square\phi = 0$, which in any lightcone set (which we denote
generically by $u, v$) reads $\partial _u \partial _v \phi = 0$. The general
solution is thus $\phi = \phi (u) + \phi (v)$ and we may restrict the analysis
to say, the $u$-part of the field only, the $v$-part being treated in an
identical way.

The Minkovski normalised wave-functions, solutions of $i \partial _U \phi _k =
k \phi_k$, are given by:

$$\eqalign{\phi _k\left( U \right)={1 \over {\left( {4\pi k}
\right)^{1/2}}}e^{-ikU}}\eqno(5)$$

\noindent
which, together with their complex conjugates, constitute a complete set if
$k \in [0, \infty]$. They define the operators $a_k$ which annihilate
the Minkovski vacuum.

$$\eqalign{a_k\equiv \left\langle {\phi _k\left| \phi  \right.} \right\rangle=
\int\limits_{-\infty }^{+\infty } {dU\phi
_k^*i\mathord{\buildrel{\lower3pt\hbox{$\scriptstyle\leftrightarrow$}}
 \over
\partial_U}\phi}}\eqno(6)$$

\noindent
whereupon

$$\eqalign{\phi \left( U \right)&=\int\limits_0^\infty  {dk\left(
{a_k\phi _k+a_k^\dagger\phi _k^*} \right)}\cr
  a_k\left|0, Mink  \right\rangle&=0\cr}\eqno(7)$$

\noindent
Similarly, the FR wave functions $\phi_{\lambda, R(L)}$ are solutions of

$$\eqalign{i\partial _{u_{R\left( L \right)}}\phi _{\lambda ,R\left( L
\right)}=ia\left|
U \right|\partial _U\phi _{\lambda ,R\left( L \right)}=\lambda \phi _{\lambda
,R\left( L \right)}}\eqno(8)$$

\noindent
and given by:

$$\eqalign{\phi _{\lambda ,R}&={1 \over {\left( {4\pi \lambda } \right)^{{1
\mathord{\left/ {\vphantom {1 2}} \right. \kern-\nulldelimiterspace}
2}}}}e^{-i\lambda u_R}\theta \left( {-U} \right)={{\theta \left( {-U} \right)}
\over {\left( {4\pi \lambda } \right)^{{1 \mathord{\left/ {\vphantom {1 2}}
\right. \kern-\nulldelimiterspace} 2}}}}\left( {-aU} \right)^{{{i\lambda }
\mathord{\left/ {\vphantom {{i\lambda } a}} \right. \kern-\nulldelimiterspace}
a}}\cr
  \phi _{\lambda ,L}&={1 \over {\left( {4\pi \lambda } \right)^{{1
\mathord{\left/ {\vphantom {1 2}} \right. \kern-\nulldelimiterspace}
2}}}}e^{-i\lambda u_L}\theta \left( {U} \right)={{\theta \left( U \right)}
\over {\left( {4\pi \lambda } \right)^{{1 \mathord{\left/ {\vphantom {1 2}}
\right. \kern-\nulldelimiterspace} 2}}}}\left( {aU} \right)^{-{{i\lambda }
\mathord{\left/ {\vphantom {{i\lambda } a}} \right. \kern-\nulldelimiterspace}
a}}\cr}\eqno(9)$$

\noindent
They are normalized and constitute another complete set if $\lambda \in
[0, \infty]$. They define the FR operators $b_{\lambda, R(L)}~~(b_{\lambda,
R(L)} = \left\langle \phi_{\lambda, R(L)} \vert \phi \right\rangle {\rm see}
(6))$.
Those operators define the FR vacuum:

$$\eqalign{b_{\lambda ,R}\left|0,R  \right\rangle\left|0,L
\right\rangle=b_{\lambda'
,L}\left| 0,R \right\rangle\left|0,L  \right\rangle=0~~\forall \lambda,
\lambda'}
\eqno(10)$$

\noindent
It is particularly useful to introduce a third complete set $(\{\phi_{\lambda,
M}\}, \lambda \in [-\infty, +\infty])$ [1, 8] which is also composed of
eigenmodes of $U\partial_U$ (see $(8)$), but built from positive energy
Minkovski modes (5) only, thus of the form

$$\eqalign{\phi _{\lambda ,M}=\int\limits_0^\infty  {dk~\alpha _{\lambda
,k}\phi
_k\left( U \right)}}\eqno(11)$$

\noindent
They are normed functions if

$$\eqalign{\alpha_{\lambda,k}={{1 \over {(2\pi k)^{1/2}}}}\left({k \over
a}\right) ^{-i \lambda /a}}\eqno(12)$$

\noindent
The associated annihilation operators are given by

$$\eqalign{a_\lambda =\int\limits_0^\infty  {dk~\alpha _{\lambda
,k}^*a_k}=\left\langle {{\phi _{\lambda ,M}}} \mathrel{\left | {\vphantom
{{\phi
_{\lambda ,M}} \phi }} \right. \kern-\nulldelimiterspace} {\phi }
\right\rangle}\eqno(13)$$

\noindent
and hence annihilate the Minkovski vacuum $\vert 0, Mink \rangle$ (7).

The Bogoliubov coefficients giving the overlap between $\phi_{\lambda, M}$ and
$\phi_{\lambda, R(L)}$, are obtained by the direct evaluation of (11) with
$\alpha_{\lambda, k}$ given by (12). One gets:

$$\eqalign{\phi _{\lambda ,M}\left( U \right)&={1 \over {\left( {8\pi ^2a}
\right)^{{1 \mathord{\left/ {\vphantom {1 2}} \right.
\kern-\nulldelimiterspace}
2}}}}\Gamma \left( {{{-i\lambda } \mathord{\left/ {\vphantom {{-i\lambda } a}}
\right. \kern-\nulldelimiterspace} a}} \right)\left[ {\theta \left( {-U}
\right)\left( {-aU} \right)^{{{i\lambda } \mathord{\left/ {\vphantom {{i\lambda
} a}} \right. \kern-\nulldelimiterspace} a}}e^{{{\pi \lambda } \mathord{\left/
{\vphantom {{\pi \lambda } {2a}}} \right. \kern-\nulldelimiterspace}
{2a}}}+\theta \left( U \right)\left( {aU} \right)^{{{i\lambda } \mathord{\left/
{\vphantom {{i\lambda } a}} \right. \kern-\nulldelimiterspace} a}}e^{-{{\pi
\lambda } \mathord{\left/ {\vphantom {{\pi \lambda } {2a}}} \right.
\kern-\nulldelimiterspace} {2a}}}} \right]\cr
&=\left[ {{{\Gamma \left( {{{-i\lambda } \mathord{\left/ {\vphantom
{{-i\lambda } a}} \right. \kern-\nulldelimiterspace} a}} \right)e^{{{\pi
\lambda
} \mathord{\left/ {\vphantom {{\pi \lambda } {2a}}} \right.
\kern-\nulldelimiterspace} {2a}}}} \over {\left( {{{2\pi a} \mathord{\left/
{\vphantom {{2\pi a} \lambda }} \right. \kern-\nulldelimiterspace} \lambda }}
\right)^{{1 \mathord{\left/ {\vphantom {1 2}} \right.
\kern-\nulldelimiterspace}
2}}}}} \right]\phi _{\lambda ,R}+\left[ {{{\Gamma \left( {{{-i\lambda }
\mathord{\left/ {\vphantom {{-i\lambda } a}} \right. \kern-\nulldelimiterspace}
a}} \right)e^{{{-\pi \lambda } \mathord{\left/ {\vphantom {{-\pi \lambda }
{2a}}} \right. \kern-\nulldelimiterspace} {2a}}}} \over {\left( {{{2\pi a}
\mathord{\left/ {\vphantom {{2\pi a} \lambda }} \right.
\kern-\nulldelimiterspace} \lambda }} \right)^{{1 \mathord{\left/ {\vphantom {1
2}} \right. \kern-\nulldelimiterspace} 2}}}}} \right]\phi _{\lambda
,L}^*\,\,\,\,\,\,\,\,\,\,\,\,\,\left( {for\  \lambda >0} \right)\cr}\eqno(14)$$

$$\eqalign{\qquad \quad \quad =\left[ {{{\Gamma \left( {{{-i\lambda }
\mathord{\left/ {\vphantom {{-i\lambda } a}} \right. \kern-\nulldelimiterspace}
a}} \right)e^{{{\pi \lambda } \mathord{\left/ {\vphantom {{\pi \lambda } {2a}}}
\right. \kern-\nulldelimiterspace} {2a}}}} \over {\left( {-{{2\pi a}
\mathord{\left/ {\vphantom {{2\pi a} \lambda }} \right.
\kern-\nulldelimiterspace} \lambda }} \right)^{{1 \mathord{\left/ {\vphantom {1
2}} \right. \kern-\nulldelimiterspace} 2}}}}} \right]\phi _{-\lambda
,R}^*+\left[ {{{\Gamma \left( {{{-i\lambda } \mathord{\left/ {\vphantom
{{-i\lambda } a}} \right. \kern-\nulldelimiterspace} a}} \right)e^{{{-\pi
\lambda } \mathord{\left/ {\vphantom {{-\pi \lambda } {2a}}} \right.
\kern-\nulldelimiterspace} {2a}}}} \over {\left( {-{{2\pi a} \mathord{\left/
{\vphantom {{2\pi a} \lambda }} \right. \kern-\nulldelimiterspace} \lambda }}
\right)^{{1 \mathord{\left/ {\vphantom {1 2}} \right.
\kern-\nulldelimiterspace}
2}}}}} \right]\phi _{-\lambda ,L}\,\,\,\,\,\,\,\,\,\,\left( {for\  \lambda
<0} \right)}\eqno(15)$$

The coefficients of the inverse transformation are readily checked to have the
same absolute value as the coefficients of the initial one as they appear in
(14) and (15). This is a simple consequence of the absence of frequency mixing
(i.e. the transformation is diagonal in $\lambda$). Thus, the relation between
$b_{\lambda, R(L)}$ and $a_\lambda, a^\dagger_\lambda$ is

$$\eqalign{b_{\lambda,R}=\alpha_\lambda a_\lambda +\beta_\lambda
a^\dagger_{-\lambda} \qquad \qquad (\lambda > 0)}\eqno(16)$$

$$\eqalign{b_{\lambda,L}=\alpha_\lambda a_{-\lambda} +\beta_\lambda
a^\dagger_\lambda \qquad \qquad (\lambda > 0)}\eqno(17)$$

\noindent
where

$$\eqalign{\alpha _\lambda =\beta _\lambda e^{{{\pi \lambda } \mathord{\left/
{\vphantom
{{\pi \lambda } a}} \right. \kern-\nulldelimiterspace} a}}=\left( {1-e^{-{{2\pi
\lambda } \mathord{\left/ {\vphantom {{2\pi \lambda } a}} \right.
\kern-\nulldelimiterspace} a}}} \right)^{-1}=\left| {{{\Gamma \left(
{-{{i\lambda } \mathord{\left/ {\vphantom {{i\lambda } a}} \right.
\kern-\nulldelimiterspace} a}} \right)e^{{{\pi \lambda } \mathord{\left/
{\vphantom {{\pi \lambda } {2a}}} \right. \kern-\nulldelimiterspace} {2a}}}}
\over {\left( {{{2\pi a} \mathord{\left/ {\vphantom {{2\pi a}}} \right.
\kern-\nulldelimiterspace} \lambda }} \right)^{{1 \mathord{\left/ {\vphantom {1
2}}
\right. \kern-\nulldelimiterspace} 2}}}}} \right|}\eqno(18)$$

\noindent
where $\alpha_\lambda$ and $\beta_\lambda$ are defined to be real by a
redefinition of the irrelevant absolute phase of the waves $\phi_{\lambda,M}$.
Hence, the unitary relation between $\vert 0, Mink \rangle$ and $\vert 0, R
\rangle \vert 0,L \rangle$ is given by [8]:

$$\eqalign{\vert 0, Mink \rangle = {\cal U} \vert 0, R \rangle
\vert 0, L \rangle = {1 \over Z} \Pi_{\lambda > 0} e^{{-e^{-{\pi \lambda /
a}}b^\dagger_{\lambda, R} b^\dagger_{\lambda,L}}} \vert 0,R \rangle \vert 0, L
\rangle}\eqno(19)$$

$$\eqalign{\vert 0, R \rangle \vert 0, L \rangle = {\cal U}^{-1} \vert 0, Mink
\rangle = {1 \over Z} \Pi_{\lambda > 0} e^{{+e^{-{\pi \lambda /
a}}a^\dagger_\lambda a^\dagger_{-\lambda}}} \vert 0,Mink \rangle}\eqno(20)$$

\noindent
where

$$\eqalign{\cal U}^{-1} = {\cal U}^\dagger = \Pi_{\lambda > 0}e^{\ah
(e^{-\pi \lambda / a})(a^\dagger_\lambda a^\dagger_{-\lambda} - a_\lambda a_{-
\lambda})}\eqno(21)$$

$$\eqalign{Z^{-1} = \Pi_{\lambda > 0} (1 / {\alpha_{\lambda}}) = \langle
Mink, 0 \vert \bigl(\vert 0, R \rangle \vert 0, L \rangle \bigr)}\eqno(22)$$

Equation (19) is the familiar rewriting of the Minkowski vacuum as a
superposition
of FR states containing excited pairs. This rewriting insures that the
expectation
value of any operator localized in the first quadrant is a thermal average.
Equation
(20) is the less familiar expression of the FR vacuum in terms of Minkowski
pair excitations
(weighted by the same factor ${{\beta _\lambda} / {\alpha _\lambda}} = e^{- \pi
\lambda / a}$). As
a corollary, the total energy of FR vacuum is greater than the energy of
Minkovski vacuum
(renormalized to zero).

In the next section we shall see how eqs. (19) and (20) encode the properties
of
the energy momentum tensor in the FR vacuum.

\medskip
\noindent
{\bf 3. The energy-momentum tensor in FR vacuum.}

The "$UU$" component of the energy-momentum tensor in FR vacuum is given by

$$\eqalign{T_{UU}^{RL}= {\langle {L,0} \vert \langle {R,0}
 \vert \left( {\partial _U\phi } \right)^2 \vert 0,R \rangle\vert 0,L
\rangle - \langle {Mink,0} \vert\left( {\partial _U\phi}
\right)^2\vert  {0, Mink}\rangle}}\eqno(23)$$
\noindent
when the negative term defines the subtraction, here purely Minkovski. Using
the
decomposition (20), and expressing $\phi$ in terms of $\phi_{\lambda,M}$ (11),
we
obtain:

$$\eqalign{T_{UU}^{RL}= 2 \int\limits_{-\infty }^{+\infty } {d\lambda \left[
{\beta
_{\left| \lambda  \right|}^2\left| {\partial _U\phi _{\lambda ,M}}
\right|^2-\alpha
_{\left| \lambda  \right|}\beta _{\left| \lambda
\right|}~Re\left( {\partial _U\phi _{\lambda ,M}\partial
_U\phi _{-\lambda ,M}} \right)} \right]}}\eqno(24)$$

\noindent
Then

$$\eqalign{\int\limits_{-\infty }^{+\infty }
{dUT_{UU}^{RL}}=~2\int\limits_{-\infty
}^{+\infty } {dU}\int\limits_{-\infty }^{+\infty } {d\lambda
}~\beta _{\left| \lambda  \right|}^2\left| {\partial _U\phi
_{\lambda ,M}} \right|^2>0}\eqno(25)$$

\noindent
because the integral of the second term of (24) vanishes due to the
orthogonality
between $\phi_{\lambda,M}$ and $\phi^*_{-\lambda,M}$. Equation (25) confirms
the
positivity of the total mean energy in FR vacuum.

Before calculation of (24), we first recall the usual expression for
$T_{UU}^{RL}$:

$$\eqalign{T_{UU}^{RL}= - {1 \over {48 \pi}} {1 \over {U^2}}}\eqno(26)$$

For purposes of comparison with (24), it is instructive to recall the two
standard derivations of (26). The first method [9] makes use of the Wightman
functions $G^+$ defined by:

$$\eqalign{G^+_{RL}(U,U')&=\langle L,0 \vert \langle R,0
\vert \phi (U) \phi(U') \vert 0,R \rangle \vert 0,L \rangle
\cr &=-{1 \over {8 \pi}} [ \theta (U) \theta (U') + \theta
(-U) \theta (-U')] ln \vert ln (U/U') \vert ^2 \cr}$$

\bigskip

$$\eqalign{G^+_M(U,U')&=\langle Mink,0 \vert \phi (U)
\phi(U') \vert 0,Mink \rangle \cr
&=-{1 \over {8 \pi}} ln \vert U - U' \vert ^2 \cr}$$

\noindent
So that

$$\eqalign{T^{RL}_{UU}&= {\displaystyle\lim_{U' \to U}}  \partial _U
\partial _{U'} [G^+_{RL}(U,U') - G^+_M(U,U')]\cr
&=(26)}\eqno(27)$$

\noindent
where the last equality is obtained by first taking the derivatives and then
taking the coincidence limit $U' \to U$ of the difference.

The second method [7] is based on Eqs. (23) and (19). The subtraction is then
the
removal of the energy of the FR quanta present in Minkovski
vacuum. Explicitely,

$$\eqalign{T_{UU} = -2 \int \limits ^\infty _0 d \lambda ~ \beta ^2
_\lambda ~\Bigl[\vert \partial_U \phi_{\lambda, R} \vert ^2 + \vert
\partial _U \phi_{\lambda, L} \vert ^2 \Bigr]}\eqno(28)$$

\noindent
The term in $\alpha_\lambda \beta_\lambda$ (see (24)) is absent because
$\left[\phi_{\lambda, R}(U) \phi_{\lambda,L}(U) =
0~~~\forall U \not=0 \right]$. One verifies that (28) gives
once more (26) because

$$\int \limits ^\infty_0 {d\lambda \over a} {\lambda \over a} \beta ^2_\lambda
= \int \limits ^\infty _0 {d\lambda \over a} {\lambda \over a} {1 \over {e ^{2
\pi \lambda / a}} - 1} = {1 \over 24}$$

The dilemma is clear; if one accepts that (26) is valid everywhere one finds
oneself in contradiction with (25).

To prove that (25) is compatible with the value of the density (26), let us go
back to (24) and evaluate $\partial _U \phi _{\lambda, M}$
with care. One has (see (11) and (12)):

$$\eqalign{i\partial _U\phi _{\lambda ,M} \equiv \int\limits_0^\infty  {{{dk}
\over {\left( {8\pi ^2} \right)^{{1 \mathord{\left/ {\vphantom {1 2}} \right.
\kern-\nulldelimiterspace} 2}}}}\left( {{k \over a}} \right)^{{{i\lambda }
\mathord{\left/ {\vphantom {{i\lambda } a}} \right. \kern-\nulldelimiterspace}
a}}e^{-ikU}}}$$

$$\eqalign{=&\mathop {\lim }\limits_{\varepsilon \to
0}{{\Gamma \left( {{{-i\lambda } \mathord{\left/ {\vphantom {{-i\lambda }
{a-1}}} \right. \kern-\nulldelimiterspace} {a-1}}} \right)} \over {\left( {8\pi
^2} \right)^{{1 \mathord{\left/ {\vphantom {1 2}} \right.
\kern-\nulldelimiterspace} 2}}}}\theta \left( U \right)\left(
{aU-ia\varepsilon } \right)^{{{-i\lambda} \mathord{\left/
{\vphantom {{-ik} {a-1}}} \right.
\kern-\nulldelimiterspace} {a-1}}}e^{{{-\pi \lambda }
\mathord{\left/ {\vphantom {{-\pi \lambda } {2a}}} \right.
\kern-\nulldelimiterspace} {2a}}}\cr
  +&\mathop {\lim }\limits_{\varepsilon \to 0}{{\Gamma
  \left( {{{-i\lambda }  \mathord{\left/ {\vphantom
{{-i\lambda } {a-1}}} \right. \kern-\nulldelimiterspace}
{a-1}}} \right)} \over {\left( {8\pi ^2} \right)^{{1
\mathord{\left/ {\vphantom {1 2}} \right.
\kern-\nulldelimiterspace} 2}}}}\theta \left( {-U}
\right)\left( {-aU+ia\varepsilon } \right)^{{{-i\lambda }
\mathord{\left/ {\vphantom {{-i\lambda } {a-1}}} \right.
\kern-\nulldelimiterspace} {a-1}}}e^{{{\pi \lambda }
\mathord{\left/ {\vphantom {{\pi \lambda } {2a}}} \right.
\kern-\nulldelimiterspace} {2a}}}\cr}\eqno(29)$$

\noindent
It is necessary to introduce the regulator $\varepsilon$ to define the
integral.
Using this expression in (25) and grouping together the contribution from
$\lambda$ and $-\lambda$ yields

$$\eqalign{T_{UU}^{RL}=\mathop {\lim }\limits_{\varepsilon \to
0}\int\limits_0^\infty  {{{d\lambda } \over {a^2}}{\lambda  \over {2\pi
}}\left[
{\beta _\lambda ^2\left( {\alpha _\lambda ^2+\beta _\lambda ^2} \right){1 \over
{U^2+\varepsilon ^2}}-2\alpha _\lambda ^2\beta _\lambda ^2\,~Re \left( {{1
\over
{U-i\varepsilon }}} \right)^2} \right]}}\eqno(30)$$

First, for $U \not= 0$, one may send $\varepsilon \to 0$, thereby recovering
(26)
because $\beta _\lambda ^2\left( {\alpha _\lambda ^2+\beta _\lambda ^2} \right)
-2\alpha _\lambda ^2\beta _\lambda ^2 = -\beta^2_\lambda$. Thus the dominant
part of the energy density comes from the interference term $\alpha _\lambda
\beta_\lambda$ in (24).

Secondly, if one integrates (30) on $U$ from $-\infty$
to $+ \infty$ one $\it{cannot}$ send $\varepsilon \to 0$ because of the
singularity at $U = 0$. So we perform the integral at fixed $\varepsilon$ and
we verify that the
$\alpha_\lambda \beta_\lambda$ term integrates to zero:

$$\int\limits^{+\infty}_{-\infty} dU Re \left({1 \over {U -
i \varepsilon}}\right)^2 = \int\limits^{+\infty}_{-\infty} dU
{{U^2 - \varepsilon^2} \over {(U^2 + \varepsilon^2)}^2} = 2 \int
\limits ^\infty _0 \partial_U \left({-U \over {U^2 +
\varepsilon^2}}\right) = 0$$

\noindent
Thus we recover the positivity of the total energy [(25)
with $\vert \partial_U \phi_{\lambda, M} \vert^2$
explicitized by the use of (29)]. Moreover, the total
energy $\int\limits^{+\infty}_{-\infty}dU T_{UU}$ diverges
as

$$\eqalign{\int\limits^{+\infty}_{-\infty} dU~T_{UU} =
\lim\limits_{\varepsilon \to 0} {1 \over \varepsilon} \left[
\int\limits^{\infty}_{0} {{d \lambda} \over {a^2}}
{{\lambda} \over {2 \pi}} \beta^2_\lambda \left(
\alpha^2_\lambda + \beta^2_\lambda \right)
\int\limits^\infty_0 dZ {2 \over {Z^2 +
1}}\right]}\eqno(31)$$

\noindent
This divergence is due to the high $k$ behaviour of
$\alpha_{\lambda, k}$ [(see (11)(12): the decreasing
character in $k$ is too weak to insure the convergence of
the total energy].

Finally let us mention that the energy content of each FR
quantum is divergent. To verify this, we compute $T_{UU}$
in the state $b^\dagger_{\lambda,R} \vert 0, Mink
\rangle$, (call it $T_{UU}^{\lambda,R}$). Using (16) and
(29), one gets

$$\eqalign{T_{UU}^{\lambda,R} &= \alpha^2_\lambda \vert
\partial_U \phi_{\lambda,M} \vert^2 \cr &=
\lim\limits_{\varepsilon \to 0} {1 \over 4\pi} \alpha
^2_\lambda {\lambda \over {a^2 (U^2 + \varepsilon^2)}}
\Bigl[\theta(U) \alpha^2_\lambda + \theta(-U)
\beta^2_\lambda \Bigr]\cr}\eqno(32)$$

\noindent
This quantity integrated over all $U$ diverges like $1
\over \varepsilon$ as well (see (31)).

The implications of those results in the case of the
accelerating detector [2, 3, 10] or the case of the
collapsing black hole [11,12,13,6] (or eternal black hole; see [15] where
static solutions which include the back reaction of a quantum field,
in the form (26), are constructed) are presently under investigation.

\bigskip

{\bf Note added in proof.} After this manuscript was completed,
our attention was called to the recent article of Unruh
[14] in which was examined the transcient behaviour of
radiation as it is put into interaction with an
accelerating detector [3]. Unruh has shown that similar
singular behaviour (on the horizon) arises, in this case
as well.

\bigskip

{\bf Acknowledgements}

\bigskip

We would like to thank Ph. Spindel for a helpful
discussion about the definition of the gamma functions and
R. Brout for interesting comments as well as a careful
reading of the manuscript. I would also thank the people
of the Racah Institute of Physics of the Hebrew University
of Jerusalem for their warm hospitality.

\vfill
\eject
\noindent
{\bf References}
\item{1.}W.G. Unruh, Phys. Rev. {\bf D14} 287 (1976).
\item{2.}W.G. Unruh and R.M. Wald, Phys Rev. {\bf D25} 942 (1982).
\item{3.}P. Grove, Class. Quantum Grav. {\bf 3} 801 (1986); D.J. Raine, D.W.
Scianna, P. Grove, Proc. R. Soc. Lond. {\bf A 435} (1991).
\item{4.}S. Fulling, Phys. Rev. {\bf D7} 2850 (1973).
\item{5.}P.C.W. Davies, J. Phys. A: Gen. Phys. {\bf 8} 609 (1975).
\item{6.}N.D. Birrel and P.C.W. Davies, {\it Quantum field in curved space},
Cambridge University Press (1982).
\item{7.}B.S. De Witt. Chap. 14 {\it in: General Relativity. An Einstein
Centenary Survey}, eds. S.W. Hawking and W. Isra\"el, Cambridge University
Press.
\item{8.}R.J. Hughes, CERN preprint TH-3670 (1983).
\item{9.}P.C. Davies and S. Fulling, Proc. R. Soc. London
{\bf A345} 59 (1977).
\item{10.}S. Massar, R. Parentani and R. Brout, ULB
preprint TH 03/92 to be published in Class. Quantum Grav.
\item{11.}S.W. Hawking `'Particle creation by black holes"
in Quantum Gravity (Calrendon Press, Oxford 1975) p. 226.
\item{12.}P.C.W. Davies, S.A. Fulling and W.G. Unruh,
Phys. Rev. {\bf D13} 2720 (1976).
\item{13.}R. Parentani and R. Brout, Int. J. of Mod. Phys.
{\bf D1} 169 (1992).
\item{14.}W.G. Unruh, Phys. Rev. {\bf D46} 3271 (1992).
\item{15.}L. Susskind and L. Thorlacius, Stanford University preprint
SU-ITP-92-12 (1992). hepth
@ xxx/9203054.

\end